# Spreadsheet Errors:
# What We Know.
# What We Think We Can Do.

**Dr. Raymond R. Panko**
**University of Hawaii**
**panko@hawaii.edu, http://panko.cba.hawaii.edu**

## Abstract

*Fifteen years of research studies have concluded unanimously that spreadsheet errors are both common and non-trivial. Now we must seek ways to reduce spreadsheet errors. Several approaches have been suggested, some of which are promising and others, while appealing because they are easy to do, are not likely to be effective. To date, only one technique, cell-by-cell code inspection, has been demonstrated to be effective. We need to conduct further research to determine the degree to which other techniques can reduce spreadsheet errors.*

## Introduction

Spreadsheets are widely used in organizations [McLean, Kappelman, & Thompson, 1993]. Each year, tens of millions of managers and professionals around the world create hundreds of millions of spreadsheets.

Although many spreadsheets are small and simple throwaway calculations, surveys have shown that many spreadsheets are quite large [Cale 1994, Cragg & King 1993, Floyd, Walls, & Marr 1995, Hall 1996]. Cragg and King [1993] audited spreadsheets as large as 10,000 cells, and when Floyd, Walls, and Marr [1995] conducted a survey of 72 end user developers in four firms, asking subjects to select a single model, the *average* model had 6,000 cells. Spreadsheets are also complex, using a large number of sophisticated functions [Hall, 1996].

Spreadsheets are also important. For instance, Gable, Yap, and Eng [1991] examined all 402 non-trivial spreadsheets in one organization. Forty-six percent were rated as important or very important to the organization, and 59% of the spreadsheets were used at least monthly. In another study, Chan and Storey [1996] surveyed 256 spreadsheet developers. Each developer was asked to describe one of their spreadsheets. When asked to identify the highest-level user of the spreadsheet's data, 13% cited a vice president, and 42% cited their chief executive officer.

Under these circumstances, if many spreadsheets contain errors, the consequences could be dire. Unfortunately, errors in bottom-line values are very likely because spreadsheet modeling is incredibly unforgiving of errors. A spelling error in a word processing document will only occasionally create a material problem; but an error almost anywhere in a spreadsheet will produce an incorrect bottom-line value. Unless the development error rate is close to one error in every ten thousand cells, *most* large spreadsheets are likely to contain errors.

In fact, we know that humans are incapable of doing complex cognitive tasks with great accuracy. The Human Error website [2000a] lists data from a number of studies of human cognitive errors. These studies indicate that even for simple cognitive tasks, such as flipping switches, error rates are about one in 200. For more complex cognitive tasks, such as writing lines of computer code, error rates are about one in 50 to one in 20.

Research on human error in many fields has shown that the problem is not sloppiness but rather fundamental limitations in human cognition [Reason 1990]. Quite simply, we do not think the way we think we think. Human cognition is built on complex mechanisms that inherently sacrifice some accuracy for speed. Speed in thinking was the overriding evolutionary force for our hunter ancestors, and although occasional errors caused the loss of hunters, new hunters were inexpensive and fun to make. Unfortunately, error rates acceptable in hunting and even in most computer applications cannot be tolerated in spreadsheets. For spreadsheet accuracy, we must somehow overcome or at least reduce human cognitive accuracy limitations dramatically.

# What We Know

Let us begin with what we actually know about spreadsheet errors and corporate practices to control spreadsheet errors. More information about the studies listed in this section is available at the Spreadsheet Research website [Panko 2000b].

## Spreadsheets Contain Errors

The Introduction noted that human error rates in complex cognitive tasks tend to be about 2% to 5%, so spreadsheet errors must be fairly frequent or we will have to throw away decades of cognitive research.

### Data from Field Studies

In fact, spreadsheet error rates actually *are* rather high. Table 1 shows data from seven field audits of real organizational spreadsheets. The field audits found errors in 24% of the 367 spreadsheets audited, and most older audits used audit techniques not likely to catch a majority of errors. The most recent field audits, in contrast, generally used better methodologies and found errors in at least 86% of the spreadsheets audited. These numbers are even more impressive when you consider that most audits only reported substantive errors, not all errors. In other words, given data from recent field audits, *most* large spreadsheets probably do contain significant errors.

## Table 1: Field Audits of Real Spreadsheets

| Study | Year | Spreadsheets | % w Errors | Cell Error Rate (CER) |
|---|---|---|---|---|
| Davies & Ikin (1) | 1987 | 19 | 21% | NR |
| Butler (2) | 1992 | 273 | 11% | NR |
| Hicks | 1995 | 1 | 100% | 1.2% |
| Coopers & Lybrand (3) | 1997 | 23 | 91% | NR |
| KPMG (4) | 1997 | 22 | 91% | NR |
| Lukasik | 1998 | 2 | 100% | 2.2%, 2.5% |
| Butler (2) | 2000 | 7 | 86% | 0.4% |
| **Overall** | **NA** | **367** | **24% (5)** | **--** |
| **1997 and Later** | **NA** | **54** | **91%** | **--** |

NR = Not reported
(1) "Serious errors"
(2) Only reported errors large enough to demand additional tax payments
(3) Spreadsheets off by at least 5%
(4) "Major errors"
(5) Weighted average

Hicks [1995] and Lukasik [1998] audited three large spreadsheets using methodologies that probably caught a large majority of errors. They found that the cell error rates, that is, the percentages of cells containing original errors (as opposed to errors based on earlier erroneous numbers in the spreadsheet) were 1.1%, 2.2%, and 2.5%. These cell error rates (CERs) are about what we would expect from general error research, as discussed above. The Butler [2000] data, in contrast, found a lower CER of only 0.38% for formula cells. This lower result may be the result of careful creation and simple accounting calculations, although the methodology used attempted to balance error detection rates and costs and used a technique that probably could not catch a majority of errors. Even the lower Butler [2000] values mean that even in spreadsheets of a few dozen cells, errors are likely.

### Data from Laboratory Experiments

As shown in Table 1, we also have data from laboratory experiments that collectively had about a thousand subjects develop spreadsheets from word problems. These word problems had known solutions, allowing 100% error detection. Across these experiments, 51% of all spreadsheets contained errors, despite the fact that most spreadsheets were only 25 to 50 cells in total size. Cell error rates for whole spreadsheet were at least 1% to 2%. These lowest CERs, furthermore, were for a task purposely designed to very simple and largely free of domain knowledge.

## Table 2: Spreadsheet Development Experiments

| Study | Year | Sample | Subjects | Spreadsheets | % w Errors | Cell Error |

| | | | | | | Rate (CER) |
|---|---|---|---|---|---|---|
| Brown & Gould | 1987 | ED | 9 | 27 | 63% | NR |
| Olson & Nilsen (1,2) | 1987-1988 | ED | 14 | 14 | NA | 21% |
| Lerch (1,2) | 1988 | ED | 21 | 21 | NA | 9.3% |
| Hassinen (2) on paper | 1988 | Ugrad | 92 | 355 | 55% | 4.3% |
| Hassinen (2) online | 1988 | Ugrad | 10 | 48 | 48% | NR |
| Janvrin & Morrison (3) Study 1, alone | 1996 | Ugrad | 78 | 61 | NR | 7% to 10% |
| Janvrin & Morrison (3) Study 1, dyads | 1996 | Ugrad | 88 | 44 | NR | 8% |
| Janvrin & Morrison (3) Study 2, alone | 1996 | Ugrad | 88 | 88 | NR | 8% to 17% |
| Kreie (post test) | 1997 | | 73 | 73 | 42% | 2.5% |
| Teo & Tan (4) | 1997 | Ugrad | 168 | 168 | 42% | 2.1% |
| Panko & Halverson, alone | 1997 | Ugrad | 42 | 42 | 79% | 5.6% |
| Panko & Halverson, dyads | 1997 | Ugrad | 46 | 23 | 78% | 3.8% |
| Panko & Halverson, tetrads | 1997 | Ugrad | 44 | 11 | 64% | 1.9% |
| Panko & Sprague (4) | 1999 | Ugrad | 102 | 102 | 35% | 2.2% |
| Panko & Sprague (4,5) | 1999 | MBA (NE) | 26 | 26 | 35% | 2.1% |
| Panko & Sprague (4,6) | 1999 | MBA (ED) | 17 | 17 | 24% | 1.1% |
| Panko & Halverson, monads | 2000 | Ugrad | 35 | 35 | 86% | 4.6% |
| Panko & Halverson, triads | 2000 | Ugrad | 45 | 15 | 27% | 1.0% |
| Total Sample | | | 998 | 1170 | 51% (7) | |

NR = not reported
ED = experienced developer
NE = not very experienced with development at work
Ugrad = undergraduate students
(1) Measured errors before subject had a chance to correct them
(2) Only measured error rate in formula cells
(3) Only measured error rate in cells linking spreadsheets
(4) Wall Task designed to be relatively simple and free of domain knowledge requirements
(5) MBA students with little or no development experience
(6) MBA students with considerable development experience
(7) Weighted average

These studies, by the way, used a variety of subjects from rank novices to highly experienced spreadsheet developers. All subject groups made errors, and when Panko and Sprague [1999] directly compared error rates for undergraduates, MBA students with little or no spreadsheet development experience, and MBA students with at least 250 hours of spreadsheet development experience, they found no significant differences in error rates across the groups.

Overall, then, intensive research has shown that spreadsheet error rates are comparable to those in other human cognitive activities. These error rates are large enough to tell us that most large spreadsheets will contain errors.

## Errors are Like Multiple Poisons

When Panko and Halverson [1997] examined errors made by their subjects, they concluded that spreadsheet errors fell into different categories. They noted that even if all errors of a certain type could be eliminated, the remaining errors would still be fatal. They compared spreadsheets errors to multiple poisons, each of which is 100% lethal.

### Quantitative and Qualitative Errors

First, there are quantitative errors and qualitative errors. A quantitative error produces and incorrect value in at least one bottom-line variable. In turn, qualitative errors, such as poor design, do not create immediate quantitative errors, but they may cause later problems in data entry or from incorrect modifications.

### Mechanical, Omission, and Logic Errors

Panko and Halverson [1997] classified quantitative errors into three basic types.

- **Mechanical errors** are simple slips, such as mistyping a number or pointing to the wrong cell when entering a formula. Mechanical errors are common in all experiments that looked at them, but the most obvious mechanical error—mistyping a number—actually has been quite rare.

- **Logic errors** occur when the developer has the wrong algorithm for a particular formula cell or expresses the algorithm incorrectly in the formula.

- **Omission errors** occur when the developer omits something that should be in the model. In experiments, this meant leaving out something in the word problem that should be in the model. In real life, it means not creating a complete model because not all factors have been considered. Non-spreadsheet research shows that humans are not good at considering all factors when considering a problem [Fischoff, Slovic & Lichtenstein 1978].

### Errors by Life Cycle

Like any other system, a spreadsheet has a life cycle, which may include needs analysis, design, construction, testing, and ongoing use. Errors may be introduced (and detected) at any stage in this life cycle.

Note that error generation does not end with the creation of the final spreadsheet. Errors may also occur during ongoing use after development. Users may enter erroneous data into the spreadsheet, of course. In addition, we will see later one particular type of post-development error, hardwiring that is distressingly common yet is also easy to prevent.

## Developers are Overconfident and Policies are Lax

Given the large amount of data on spreadsheet errors, one might think that companies would implement strong policies for spreadsheet development and testing. However, that rarely is not the case.

Most studies that have audited spreadsheets or surveyed users have reported poor development practices [Cragg & King 1993, Davies & Ikin 1987, Hall 1996, Schultheis & Sumner 1994]. In addition, studies that looked at corporate controls also found a general pattern of little control and of the controls that did exist being largely informal [Cale 1994, Cragg & King 1993, Davies & Ikin 1987, Floyd, Walls, & Marr 1995, Galletta & Hufnagel 1992, Hall 1996, Speier & Brown 1996].

One reason for this lack of disciplined development practices and policies may be that spreadsheet developers are overconfident in the accuracy of their spreadsheets. Certainly laboratory studies, field audits, and surveys have shown a high degree of confidence in the accuracy of spreadsheets, even if quite a few errors were found later [Brown & Gould 1987, Davies & Ikin 1987, Floyd, Walls, & Marr 1995, Panko & Halverson 1997, Panko & Featherman 1999].

In one experiment [Panko & Featherman 1999], for instance, developers were asked to estimate the likelihood that they had made an error during development. The median estimate was 10%, and the mean was 18%. In fact, 86% had made an error in their spreadsheet. When debriefed in class and asked to raise their hands if they thought they were among the successful 14%, well over half of all subjects raised their hands. Although these subjects were students, overconfidence has also been found among experts in many fields [Johnson 1988, Shanteau 1992]. Indeed, overconfidence is one of the most consistent findings in behavioral sciences [Plous 1993] and has been linked to a lack of care in risk avoidance [Rasmussen 1990, Rumar 1990].

## What We Think We Can Do

Now that the widespread existence of errors in spreadsheets is well documented, the next logical question is, "What can we do to reduce errors?" Notice that the word is "reduce" rather than "eliminate." Years of human error research have shown that there is simply no way to eliminate error completely.

### The Tenacity of Error

Human error research indicates that human error is tenacious because people are not terribly good at detecting and correcting errors. The Human Error website [Panko 2000a] shows that error detection and correction rates approaching 90% only occur in the simplest processes, such as proofreading for spelling errors in which the misspelling is not itself a valid word. If the result of the spelling error is itself a valid word, error detection rates fall to about 70%. Even this is high compared to error detection and correction for logical errors in mathematics, which in Allwood's [1984] classic study succeeded in only about half of all errors. Error detection and correction for omission errors is much lower still [Allwood 1984, Bagnara, *et al.* 1987, Woods 1984].

More directly, the Human Error website [Panko 2000a] has data from a number of software team code inspection studies, in which a *group* of programmers goes over a program one line at a time to look for errors. These intensive code inspections only find around 80% of all errors despite the use of team code inspection by programming professionals.

Even more directly, there have been experiments in spreadsheet code inspection, in which subjects examine a spreadsheet cell-by-cell to look for errors. These studies indicate that subjects working alone only catch about half of all errors [Galletta *et al.* 1993 1997, Panko 1999], even when the subjects are experienced spreadsheet developers [Galletta *et al.* 1993 1997]. We also know that error detection rates in spreadsheet code inspections only approach 90% for mechanical typing and pointing errors in short formulas [Panko 1999]. For logic errors, mechanical errors in long formulas, and omission errors, detection rates in spreadsheet code inspection are far lower [Panko 1999].

Software developers, who are highly experienced with errors, respond to the difficulty of detecting errors by engaging in massive amounts of formal testing. About a third of the total software development effort goes into formal testing [Grady 1994], and even after several stages of testing, errors remain in about 0.1% to 0.3% of all lines of code [Putnam & Myers 1992].

Given the tenacity of error in the face of intensive code inspection and other types of testing, we should not be very sanguine about any technique of error reduction that falls short of massive testing.

## Cell Protection

However, there appears to be one simple thing we can do to prevent at least one type of error. Earlier, we noted that one type of error during ongoing use, hardwiring, is very common. In hardwiring, a user cursors to a formula cell and enters a number in the cell. This generally occurs because the user did not realize that the cell was a formula cell and thought they should enter the value. One survey of an Australian mining company's spreadsheets found hardwiring errors in about 30% of the spreadsheets [Dent 1995].

When a spreadsheet is hardwired, it is likely to be correct for the current user, provided the correct current value is entered into the formula cell. However, if the spreadsheet is saved and then run again, the spreadsheet will not be correct for subsequent users.

Hardwiring errors are rather easy to prevent using cell protection. Cell protection only allows users to change pre-specified input cells, so that if a user attempts to hardwire a formula cell, he or she will be prevented from doing so. Unfortunately, although cell protection is fairly effective and easy to do, it is only done in half or fewer of all spreadsheets [Cragg and King 1993, Davies & Ikin 1987, Hall 1996].

## Re-Keying Data

There is also a simple way to reduce data input errors. This is to have two input sections, so that all data will be entered twice. It is rather trivial to highlight differences between two blocks of input data in order to highlight errors. This method, used in traditional data processing, is called verification. Data entry errors are likely to be random, so making the same error twice is unlikely. It is easy to check if two input areas are the same and, if not, to determine where the error lies.

## Examining Results for Reasonableness

Ineffective home remedies in the United States are called "whiskey cures," after the old aphorism, "Of all the things that do not cure the common cold, whiskey is the most popular." With a whiskey cure, you take a step to reduce harm, but the step is largely or entirely ineffective.

One commonly seen activity in spreadsheet development is looking over a spreadsheet's results for reasonableness [Hendry & Green 1994, Nardi & Miller 1991]. If a few errors are found, the seeker feels that he or she is very effective at finding errors. If no errors are found, then the seeker feels that the spreadsheet is free of error.

Given the difficulty of finding errors noted above even when full cell-by-cell code inspection is used, merely looking over a spreadsheet's numbers for reasonableness must be considered a whiskey cure. In addition, studies [Klein 1997, Rickets 1990] have shown that people are not very good at finding errors when they assess numbers for correctness. Although looking over results for reasonableness is simple and inexpensive and should be done, it must not be considered an acceptable stopping point.

## Good Development Practices?

A number of authors have proposed good spreadsheet development practices [e.g., Kreie 1988, Rajalingham, *et al.* 2000, Ronen, Palley, & Lucas, 1989], often modeled after good software development practices. For instance, they usually advocate a full systems development life cycle, with definite needs analysis and design stages, both of which often are skipped by spreadsheet developers [Brown & Gould 1987, Hall 1996]. They usually also propose modular design and the placement of all input numbers in a single "data" or "assumptions" section. Finally, a few propose something like clean room development, in which equations are worked out ahead of time [Rajalingham, *et al.* 2000], although they usually stop short of formal proofs.

One general problem with these methodologies is that they have not been tested experimentally to see if they really reduce errors and, if so, by how much. In the terminology of the U.S. Federal Drug Administration, they have not been proven "safe and effective." Experiments conducted to reduce errors [e.g., Janvrin & Morrison 1996, Kreie 1988] have found it very difficult to create statistically significant reductions in error rates. In an unpublished study by the author, when spreadsheets using assumptions sections and not using assumptions sections were compared, there was almost no difference in error rates. Indeed, traditional "good practice" actually may be harmful. If all input data are placed in one section and logic in another, the logic may be more difficult to read for code inspection.

Overall, although good practice in development probably is commendable, it must not be considered sufficient unless it is proven safe and effective through experimentation. Certainly, we should not listen to claims of effectiveness in this area until methods have been tested.

### Error Discovery Software

Another promising but untested area is error discovery software designed to identify errors in spreadsheets. Error discovery software comes in two basic forms. One form graphically portrays patterns of connections between cells within spreadsheets, say by coloring or arrows, to highlight patterns that may indicate possible errors in the spreadsheet. However, it is not clear how many errors seen in experiments and field audits today would be made more visible by this type of software [Panko, 2000c]. Again, there is a need for testing before claiming that such tools safe and effective.

The other type of error discovery software acts like a spell checker or grammar checker, highlighting specific cells that may contain errors. Again, however, we would like to know the extent to which such software could actually catch the types of errors found in experiments and field audits.

Another issue is customization to specific purposes. The Butler [2000] software, for instance, has several error-detection techniques tailored to the ways in which people may be deliberately cheating on taxes, for instance making a number in a column a label and then right-justifying it, to reduce the column total.

### Code Inspection

In software development, about a third of all development time is spent in formal testing, as noted earlier. Testing is considered the sine qua non for reducing errors in software development. Even when good practice is used throughout the needs assessment, design, and code development process, there is no substitute for testing.

In spreadsheet development, testing is fairly rare [Cragg & King 1993, Gable, Yap & Eng 1991; Hall 1996, Nardi 1993, Schultheis & Sumner 1994]. Informal comments often cite the cost of testing as being prohibitive. However, without formal testing, the cost of errors must be considered.

There are two general forms of testing. One is execution testing, in which known data are used to test the spreadsheet. However, known data are not always available because spreadsheets programs often allow far more complex analyses than a company could previously perform. In addition, execution testing is a complex craft in which out-of-bounds and extreme values must be tested, and in which test cases generally must be selected very carefully. Yet even when execution testing is done with spreadsheets, it typically avoids such niceties [Hall 1996].

The other form of testing is code inspection, in which an individual or group goes through a spreadsheet cell by cell to look for errors. Code inspection is very expensive, because software development has taught us that the yield (percentage of errors found) in software testing falls when code inspection is done rapidly [Panko 2000a]. In addition, code inspection is exhausting work, and inspectors tend to hate it. Finally, as noted above, individual code inspectors find only half or fewer of all errors, so group code inspection is necessary. Overall, group code inspection is an unpleasant medicine yet the only medicine that has been experimentally tested to date and found to be fairly effective, catching about 80% of all errors [Panko 2000a].

At the same time, there can be cost-yield tradeoffs. For instance, Butler [1992 2000] was concerned with finding incorrect spreadsheets in tax auditing. Each error found might result in income from additional taxation. However, auditing is also expensive. Given tradeoffs between yield and cost, it was reasonable to use single-person code inspection augmented by software, and was reasonable not to put limits on inspection rates. In general, the Butler [2000] analysis method represents an interesting approach for matching costs with yields.

However, it is important for firms to deliberately consider how to do code inspection. If accuracy is important, single-person code inspection appears to be a poor idea. Teams of three or more are likely to be necessary.

## Conclusion

Research on spreadsheet errors began over fifteen years ago. During that time, there has been ample evidence demonstrating that spreadsheet errors are common and nontrivial. Quite simply, spreadsheet error rates are comparable to error rates in other human cognitive activities and are caused by fundamental limitations in human cognition, not mere sloppiness. Nor does ordinary "being careful" eliminate errors or reduce them to acceptable levels.

Now that the reality of spreadsheet error has been established, the next step is to ask what we can do to reduce spreadsheet errors. Unfortunately, only one approach to error reduction has been demonstrated to be effective. This is code inspection, in which a *group* of spreadsheet developers checks a spreadsheet cell-by-cell to discover errors. Even this exhausting and expensive process will catch only about 80% of all errors. Other things can be done to reduce errors, but given the tenacity of spreadsheet error in the face of even cell-by-cell code inspection, we should be careful about expecting too much from most error-reducing approach. What is needed now is an experimental research program to determine which approaches really are safe and effective.